\documentclass{aa}  

\usepackage{graphicx}
\usepackage{epstopdf}
\usepackage{txfonts}
\usepackage{lipsum}
\usepackage{subcaption}         
\usepackage{lscape}             
\usepackage{placeins}           
\usepackage{hyperref}           
\usepackage{verbatim}
\usepackage{siunitx} 
\usepackage{float}

\begin{document}

   \title{Zeeman-Doppler imaging of V889 Her}

   \subtitle{Surface spots and magnetic field topology 2017 -- 2023}

\author{T. Hackman
          \inst{1}
          \and
          J.J. Lehtinen\inst{1,2}
           \and
          O. Kochukhov\inst{3}
         \and
         P. Pirttikangas \inst{1}
          \and I. Ilyin \inst{4}
          \and A. Haris \inst{1}
          \and M.J. Korpi-Lagg \inst{5}
          \and S. Wedemeyer \inst{6}
          \and M. Tuomi \inst{1}
        }
    
          \institute{Department of Physics, P.O. Box 64, FI-00014 University of Helsinki, Finland \\ \email{thomas.hackman@helsinki.fi}
          \and
          Alshain Oy, Jaakkimantie 24A, FI-02140, Espoo, Finland
\and
Department of Physics and Astronomy, Uppsala University, Box 524, S-75120 Uppsala, Sweden
\and
Leibniz Institute for Astrophysics Potsdam (AIP), An der Sternwarte 16, 
D-14482 Potsdam, Germany
\and 
Department of Computer Science, Aalto University, PO Box 15400, FI-00076 Espoo, Finland
\and Rosseland Centre for Solar Physics, University of Oslo, PO Box 1029, Blindern, N-0315 Oslo, Norway
}

   \date{Received, 20XX}

 
  \abstract
   {The surface magnetic field of the young solar analogue \object{V889 Her} is studied using Zeeman-Doppler imaging (ZDI). This well-studied star has previously been reported to have an exceptionally high differential rotation and show abrupt changes in the surface magnetic field and spot configuration.}
   {Our aim is to study the evolution of the magnetic field topology in
     V889 Her. This will contribute to understanding dynamo action in young solar-type stars. We also want to constrain the surface differential rotation curve of the target star.}
 {We employ the ZDI method on observations collected with the SOFIN spectropolarimeter mounted at the Nordic Optical Telescope. We test differential rotation parameters previously reported in studies of the star.}
 {We observe significant changes in the radial field near the rotational pole  throughout the whole span of the observations. Specifically, the polarity of the radial field is reversed around 2021. The large-scale magnetic field also contains a persistent strong azimuthal component at high latitudes. We detect abrupt changes in the spot configuration in June 2018 on the time scale of a week. The observations favour a solar type differential rotation. The different rotation parameters influence the brightness map more than the magnetic field topology.}
   {We cannot link the observed changes in the magnetic field topology with the evolution of the long-term spot activity. The abrupt changes observed in 2018 confirm previous reports that the surface magnetic field and spot configuration may undergo rapid evolution on short time scales. This complicates the estimation of the surface differential rotation.}

   \keywords{stars: magnetic field -- stars: activity -- stars: solar-type -- stars: individual: V889 Her -- starspots
               }
\maketitle
\nolinenumbers

\section{Introduction}

The solar analogue \object{V889 Her} (HD 171488) is an extremely active BY Dra star of spectral type G2V with an estimated age of 30-50 Myr \citep{Kazarovets1997,Strassmeier2003}. 
It has been the target of several extended studies that have reported strong differential rotation and rapid changes in its spot activity \citep[see, e.g.,][]{Willamo2019, Willamo2022, Brown2024}. Attempts to identify regularities in
its magnetic activity have so far been inconclusive. Furthermore, its
surface differential rotation has been a major problem to determine, as it may even be significantly varying  \citep[see, e.g., Fig. 2 in][]{Brown2024} and could be non-monotonic \citep{Tuomi2024}. This makes V889 Her a challenging but interesting target for monitoring.

In this study, we used data collected with the SOFIN spectropolarimeter mounted on the 2.56 m Nordic Optical Telescope \citep{Ilyin2000}. The observations span the years 2017--23 (2020 and 2022 missing). Thus, our observations  partly overlap the long-term study by \cite{Brown2024}. The observations were used to calculate Zeeman-Doppler imaging (ZDI) maps of both the surface magnetic field vector and brightness distribution. For this we used the \texttt{inversLSD2} code developed by \cite{Kochukhov2026}. The differential rotation of the star was implemented using three different sets of parameters: No significant differential rotation \citep{Willamo2019}, monotonic differential rotation \citep{Brown2024}, and non-monotonic differential rotation \cite{Tuomi2024}. The aim of this comparison was to evaluate which case of differential rotation would give the best fit to the observations. 

The results were used to study the evolution of surface spots and magnetic fields, in particular possible polarity reversals. In the Sun, the magnetic field reversal can be seen first in the radial field surrounding the rotation poles and later in the azimuthal field near the solar equator. However, more active solar type stars have not in general been found to "behave" in the same way, although several cases of polarity reversals have been reported \citep[see, e.g.,][and references therein]{Willamo2022a}.
In addition to comparing ZDI maps for different years, we also explored the possibility of fast changes occurring during the observation runs. 

\section{Observations}

\begin{figure}
    \centering
    \includegraphics[width=0.8\linewidth, angle=90]{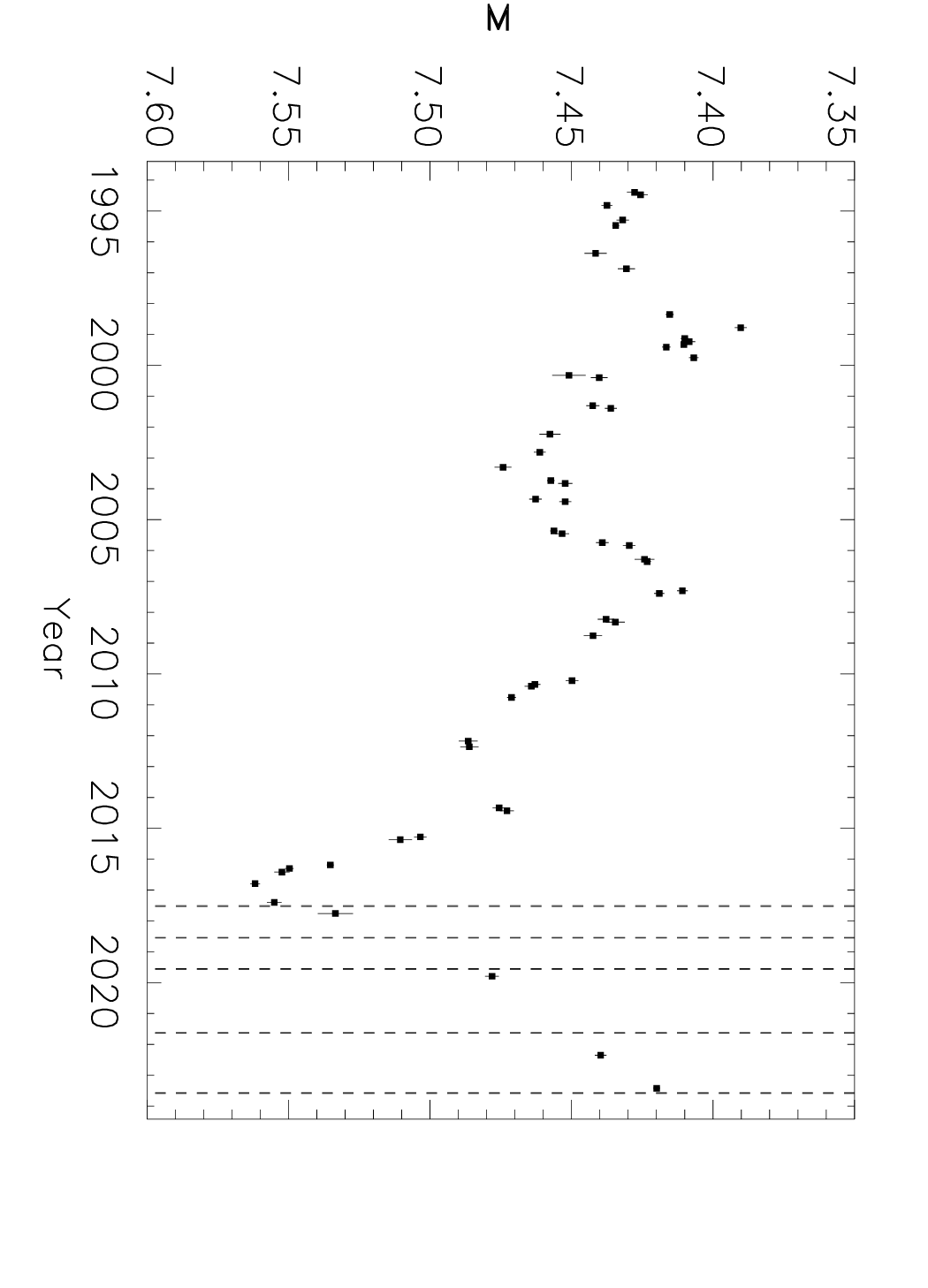}
    \caption{Long-term mean V-magnitudes ($M$) of V889 Her. The vertical dashed lines mark the timings of the SOFIN observations.}
    \label{photom}
\end{figure}

The spectropolarimetric observations were obtained in June--July 2017, 2018, 2019, 2021, and 2023. The observations are summarized in Table \ref{obssum} and a complete log of these can be found in Table \ref{obs}.
The timing of the observations compared to the long-term mean V-magnitudes of the V889 Her is shown in Fig. \ref{photom}. The photometry is the same that was used in the study by \cite{Tuomi2024}. The independent mean magnitudes were derived using the CPS method applying a 30 day window to the data. This method is described in \cite{Lehtinen2011}. As can be seen, all observations occurred during a long-term brightening of the star, i.e. a period of gradually declining spot activity after the strongest maximum recorded in long-term photometry.

The observations spanned 7--10 nights, and 6--12 observations ($n_\phi$ in Table  \ref{obssum}) were collected depending mainly on the weather conditions. The typical $S/N$ range was 250--350 measured at $\lambda = 6300$ \AA. The observations followed the usual procedure for SOFIN described by \cite{Ilyin2012}. The basic reduction was done using the  package described in \cite{Willamo2019} based on the 4A software \citep{Ilyin2000}. The exposure times are measured from the start of the first to the end of the last sub-exposure. This time was typically $~ 50$ min, which would mean a phase smearing of $\delta \phi \approx 0.026$.

\begin{table}
  \caption{Summary of the SOFIN observations 2017 -- 2023.}
  \centering
\begin{tabular}{cccrc}
  \hline \hline

Set & Dates & BJD$_0$ - 2400000 & $n_\phi$ & $n_{\rm LSD}$ \\ 
\hline
2017 & 04/06 -- 14/06/2017 & 57903.686 &  6 & 1817\\ 
2018 & 15/06 -- 23/06/2018 & 58280.013 & 9 & 1101 \\ 
2019 & 20/06 -- 29/06/2019 & 58651.367 & 12 & 1890 \\ 
2021 & 17/07 -- 23/07/2021 & 59408.783 & 9 & 1966\\ 
2023 & 27/06 -- 07/07/2023 & 60119.372 & 11 & 1726\\ 
\hline
\end{tabular}
\label{obssum}
\end{table}

Since the S/N-ratio of single spectral lines is far from sufficient to detect a signal in Stokes V, least-squares-deconvolved (LSD) spectra were computed using an IDL interface to the {\verb|iLSD|} code developed by \cite{Kochukhov2010}. We refer to that paper and to \cite{Donati1997} for the discussion of general principles and computational details of the method.
The line
mask for the LSD calculations was extracted from the Vienna Atomic Line Database \citep[VALD,][]{VALD1,VALD2}.

The resulting Stokes V profiles are shown in Fig. B.1 of Appendix \ref{appB}.
The numbers of spectral absorption lines included in the LSD profiles $n_{\rm LSD}$ are listed in Table \ref{obssum}.
This resulted in mean Stokes I and V spectra, with a $S/N$ of $\sim 5000$. The mean wavelength of the LSD profiles was $\lambda_{\rm LSD} \approx 5000$ {\AA} and the effective Land\'e factor $g_{\rm eff} \approx 1.20$.

Contrary to what usually is the practice, we do not present any null polarisation profiles. The reason is that the SOFIN instrument is mounted at Cassegrain focus and not fibre fed. This causes a reduced wavelength stability, which may distort the null profile. This is not a problem for the Stokes V-profiles, as these are derived from pairs of right- and left-handed circular polarisation signals obtained in the same CCD-frame. But when obtaining the null profiles, one has to combine different CCD-frames, where the pairs are switched.

\begin{table}
  \caption{The three sets of stellar parameters used for deriving ZDI maps.}
  \centering
\begin{tabular}{llllll}
  \hline \hline

  & $P_{\rm eq}$ [d]  & $a$ & $b$ & Reference \\
\hline
W19       & 1.33697             & 0      & 0 & \cite{Willamo2019} \\
B24       & 1.328                & -0.0674 & 0 & \cite{Brown2024} \\
T24       & 1.362               & 0.130 & -0.173 & \cite{Tuomi2024}\\
\hline
\end{tabular}
\label{spar}
\end{table}

\begin{figure*}
    
   \centering

\includegraphics[width=1.0\textwidth]{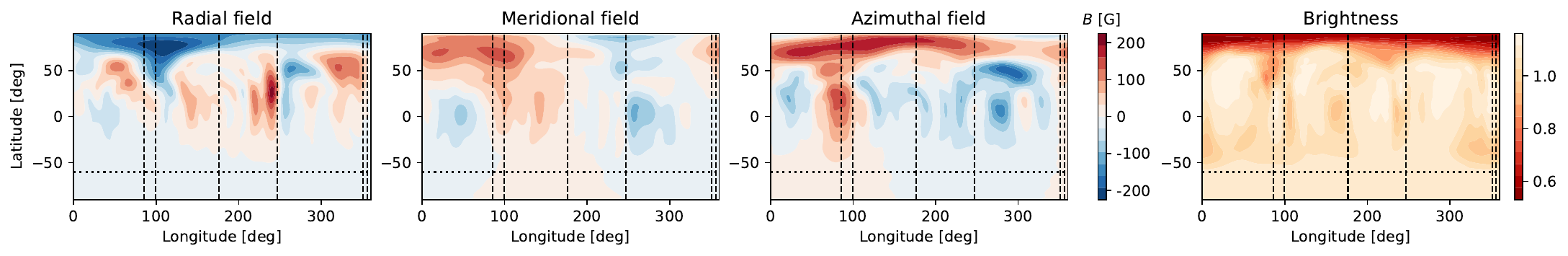}

\includegraphics[width=1.0\textwidth]{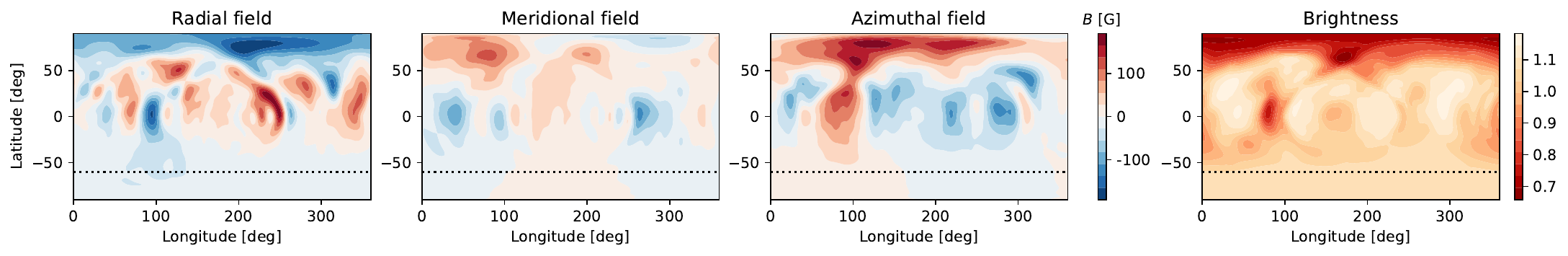}

\includegraphics[width=1.0\textwidth]{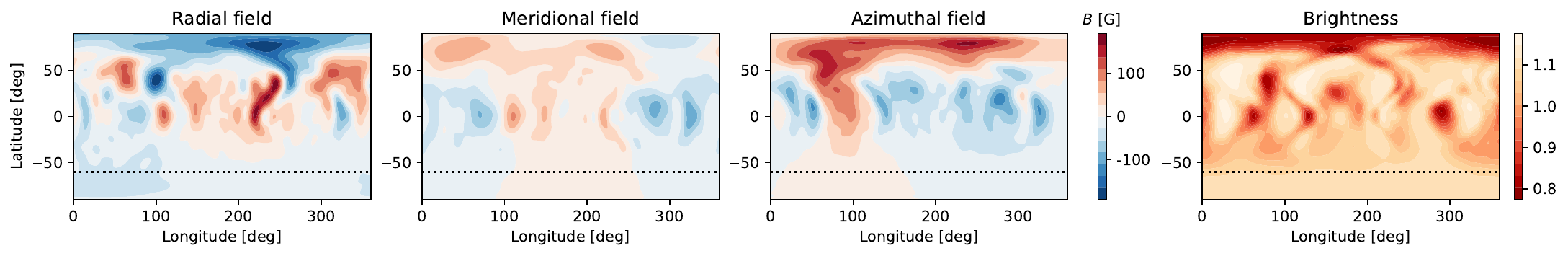}

\caption{Resulting ZDI maps from 2017 using the three different sets of rotation parameters: From top to bottom W19, B24 and T24. The brightness corresponds to the relative intensity at the wavelength of the LSD profiles. The vertical lines indicate the observed rotation phases and the horizontal lines show the limit of the visible hemisphere.}
              \label{ZDI2017}
\end{figure*}

\begin{figure*}
    
   \centering

\includegraphics[width=1.0\textwidth]{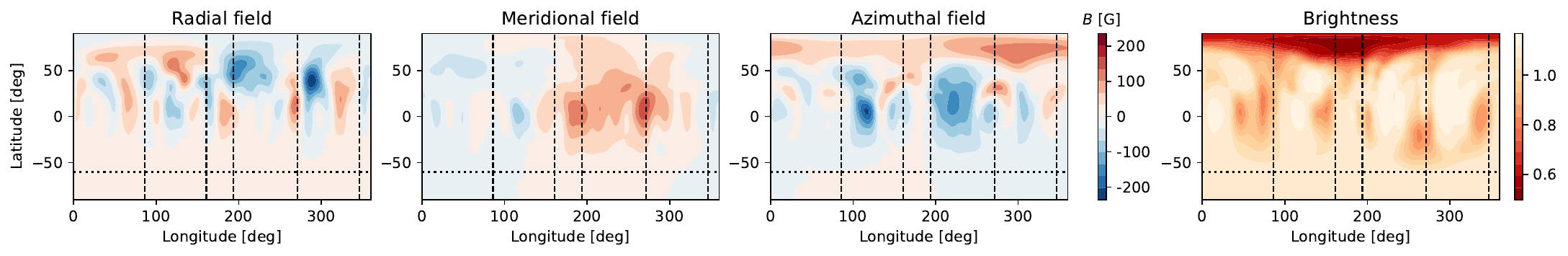}

\includegraphics[width=1.0\textwidth]{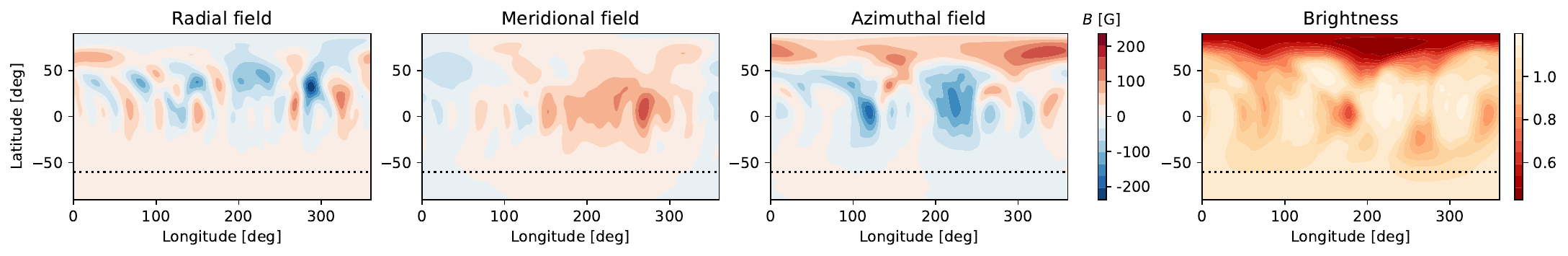}

\includegraphics[width=1.0\textwidth]{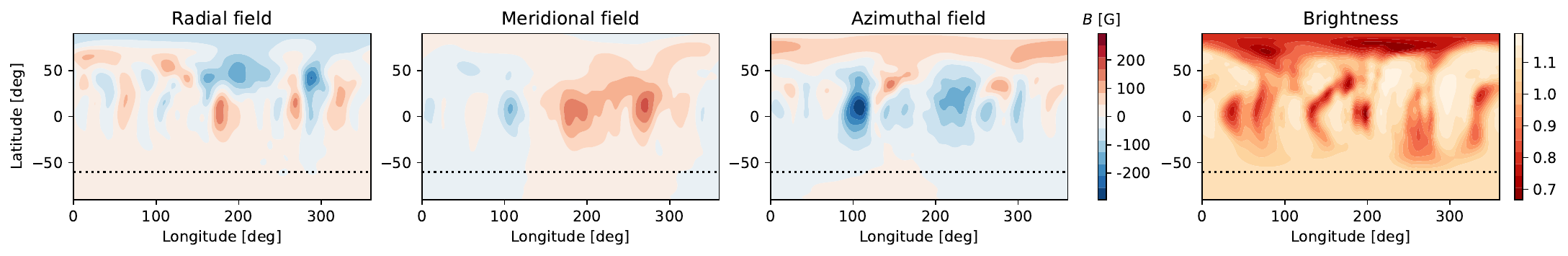}

\caption{Same as Fig. \ref{ZDI2017} but for the first five observations from 2018.}
              \label{ZDI2018f5}
\end{figure*}

\begin{figure*}
    
   \centering

\includegraphics[width=1.0\textwidth]{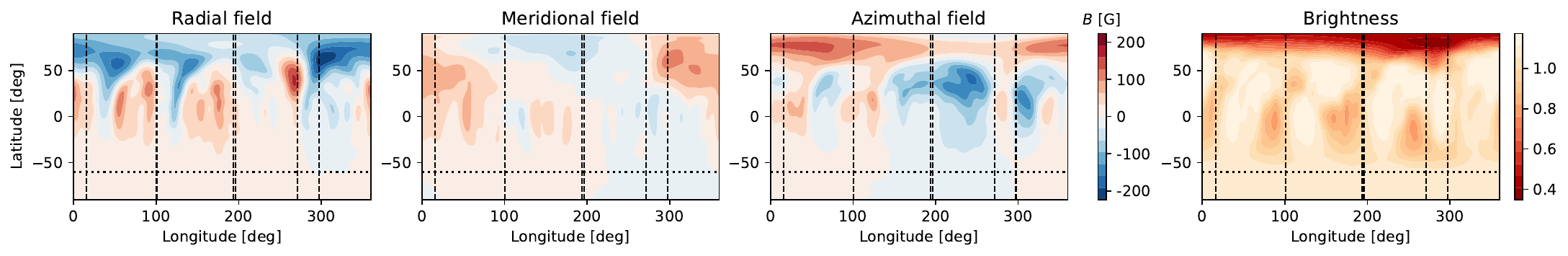}

\includegraphics[width=1.0\textwidth]{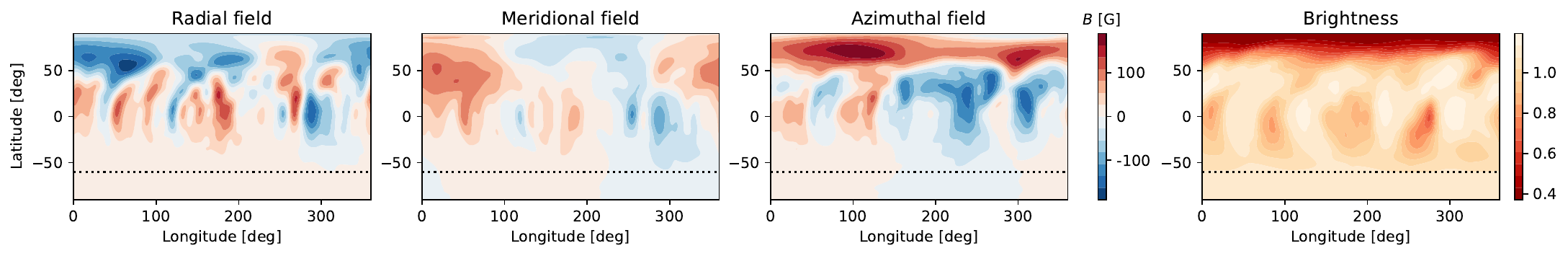}

\includegraphics[width=1.0\textwidth]{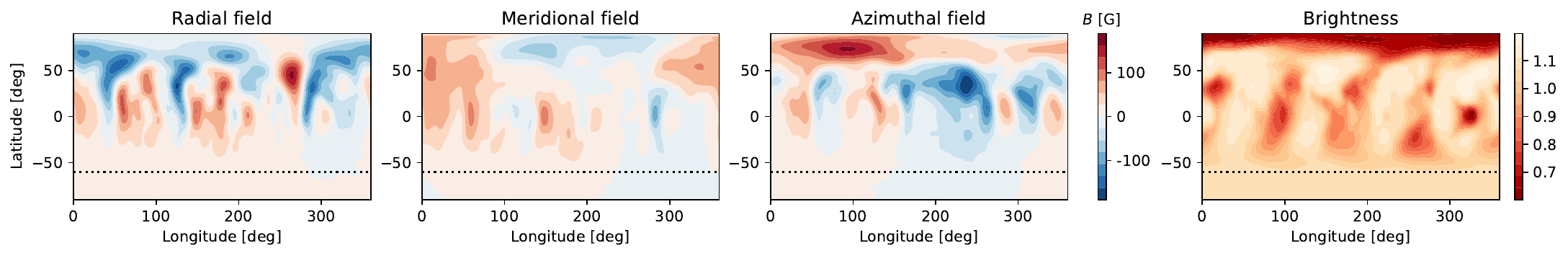}

\caption{Same as Fig. \ref{ZDI2017} but for the last six observations from 2018.}
              \label{ZDI2018l6}
\end{figure*}

\begin{figure*}
    
   \centering

\includegraphics[width=1.0\textwidth]{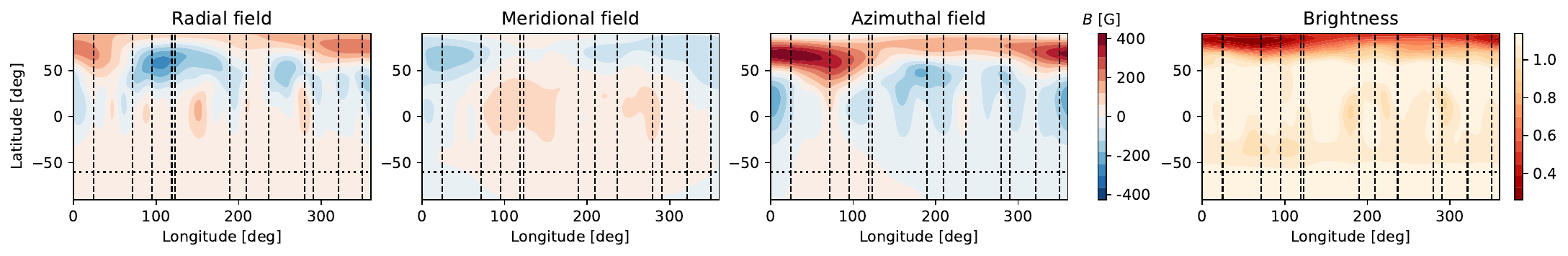}

\includegraphics[width=1.0\textwidth]{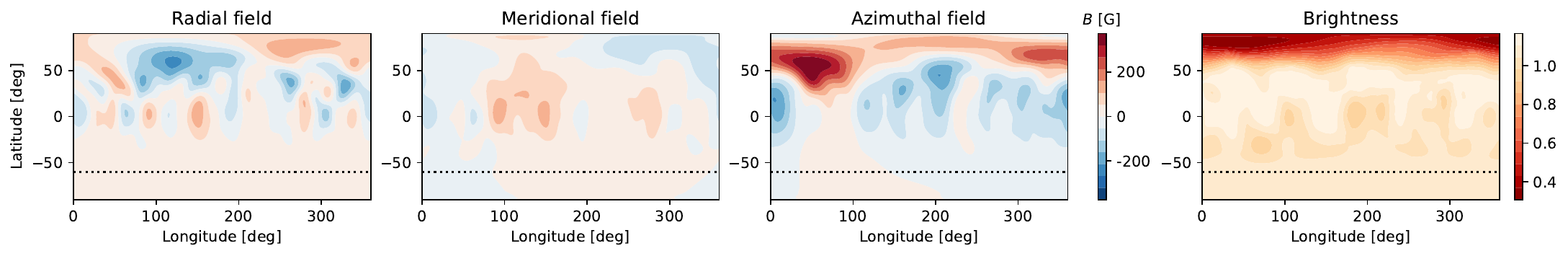}

\includegraphics[width=1.0\textwidth]{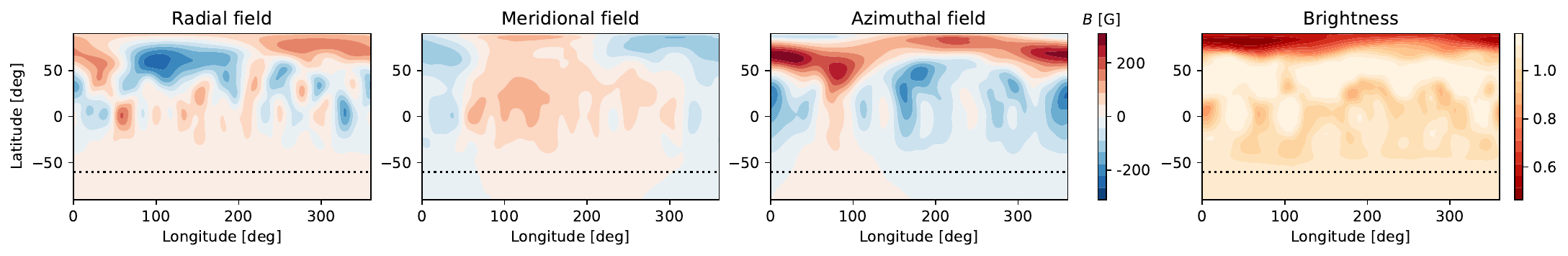}

\caption{Same as Fig. \ref{ZDI2017} but for the  observations from 2019.}
              \label{ZDI2019}
\end{figure*}

\begin{figure*}
    
   \centering

\includegraphics[width=1.0\textwidth]{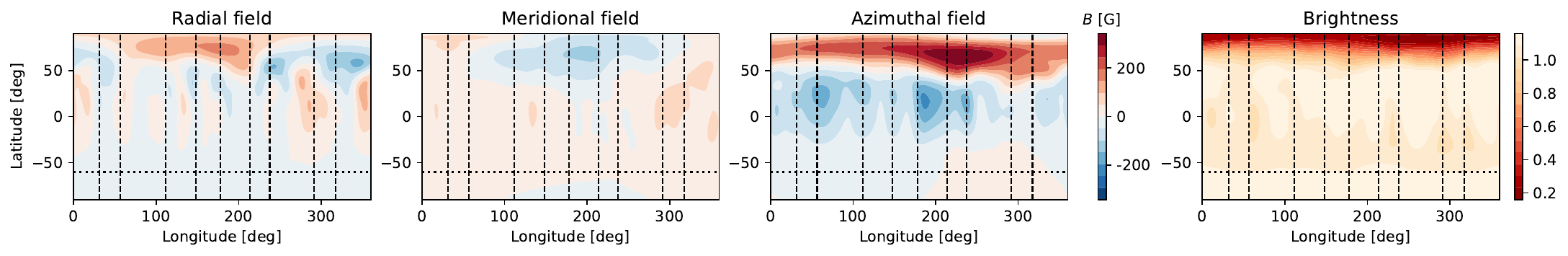}

\includegraphics[width=1.0\textwidth]{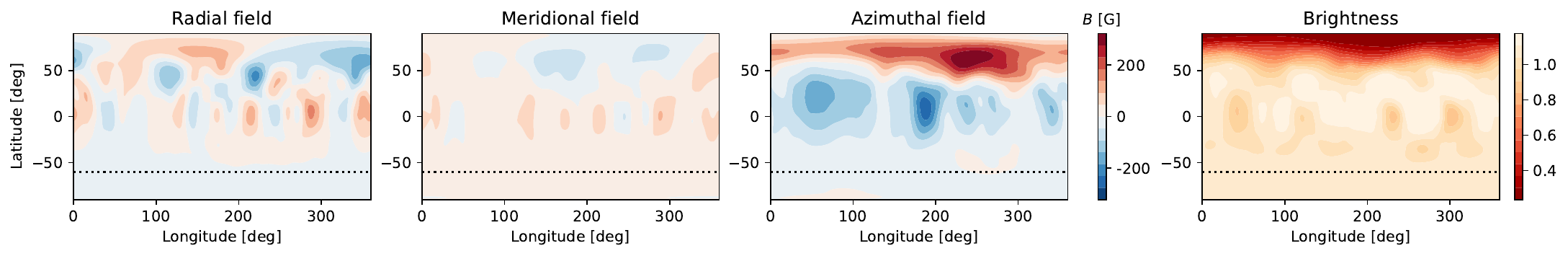}

\includegraphics[width=1.0\textwidth]{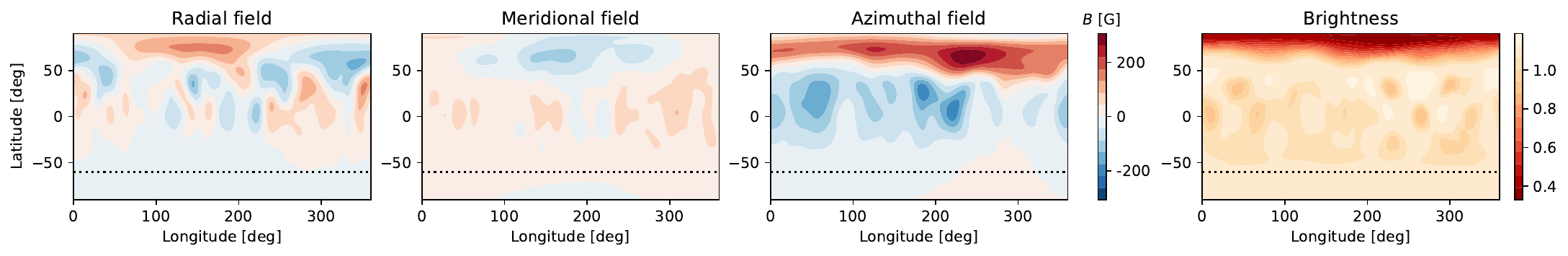}

\caption{Same as Fig. \ref{ZDI2017} but for the  observations from 2021.}
              \label{ZDI2021}
\end{figure*}

\begin{figure*}
    
   \centering

\includegraphics[width=1.0\textwidth]{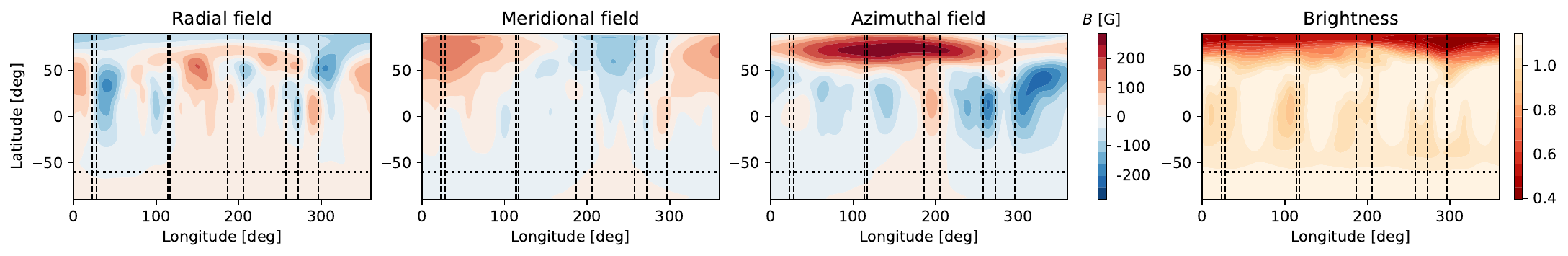}

\includegraphics[width=1.0\textwidth]{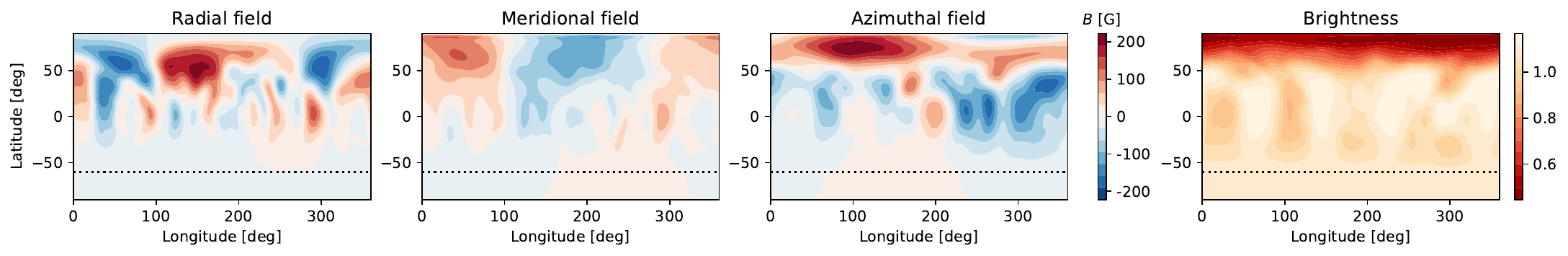}

\includegraphics[width=1.0\textwidth]{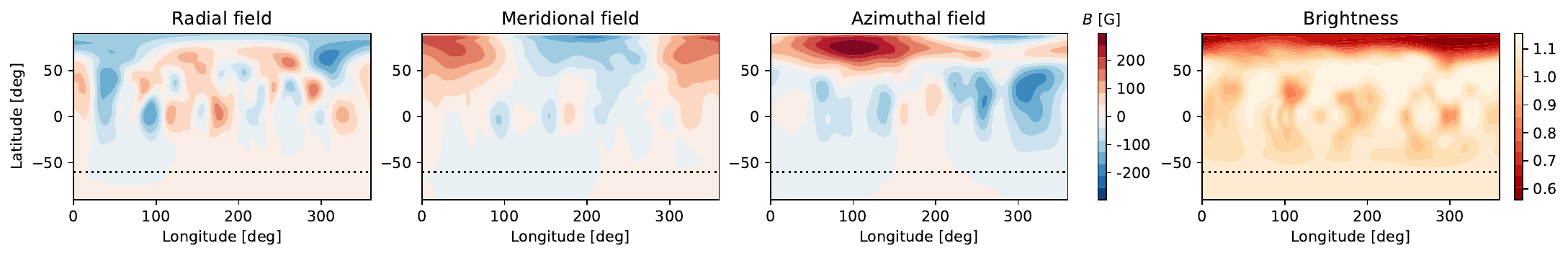}

\caption{Same as Fig. \ref{ZDI2017} but for the  observations from 2023.}
              \label{ZDI2023}
\end{figure*}

\section{Zeeman-Doppler imaging method}
\label{ZDI}

The Stokes I and V LSD profiles were used to calculate surface magnetic field and
brightness maps using the {\verb|inversLSD2|} code 
\citep{Kochukhov2026}. This code is a new extended version of {\verb|inversLSD|} \citep{Kochukhov2014}.
The magnetic field 
geometry in this code is based on spherical harmonics 
\citep[see, e.g.,][]{Kochukhov2014}. The geometric resolution is limited by setting a maximum angular degree 
$\ell_{\max}$ for the solution. We used $\ell_{\max} = 20$, justified by the simulations of \cite{Hackman2024}. The reconstruction of the surface brightness is based on a geometry, where the surface is divided into approximately equal sized elements, meaning a gradually increasing number from the poles to the equator. The total number of elements used for \object{V889 Her} was 1876.

\begin{figure*}
   \vspace{0.5cm}
 \includegraphics[width=7.5cm]{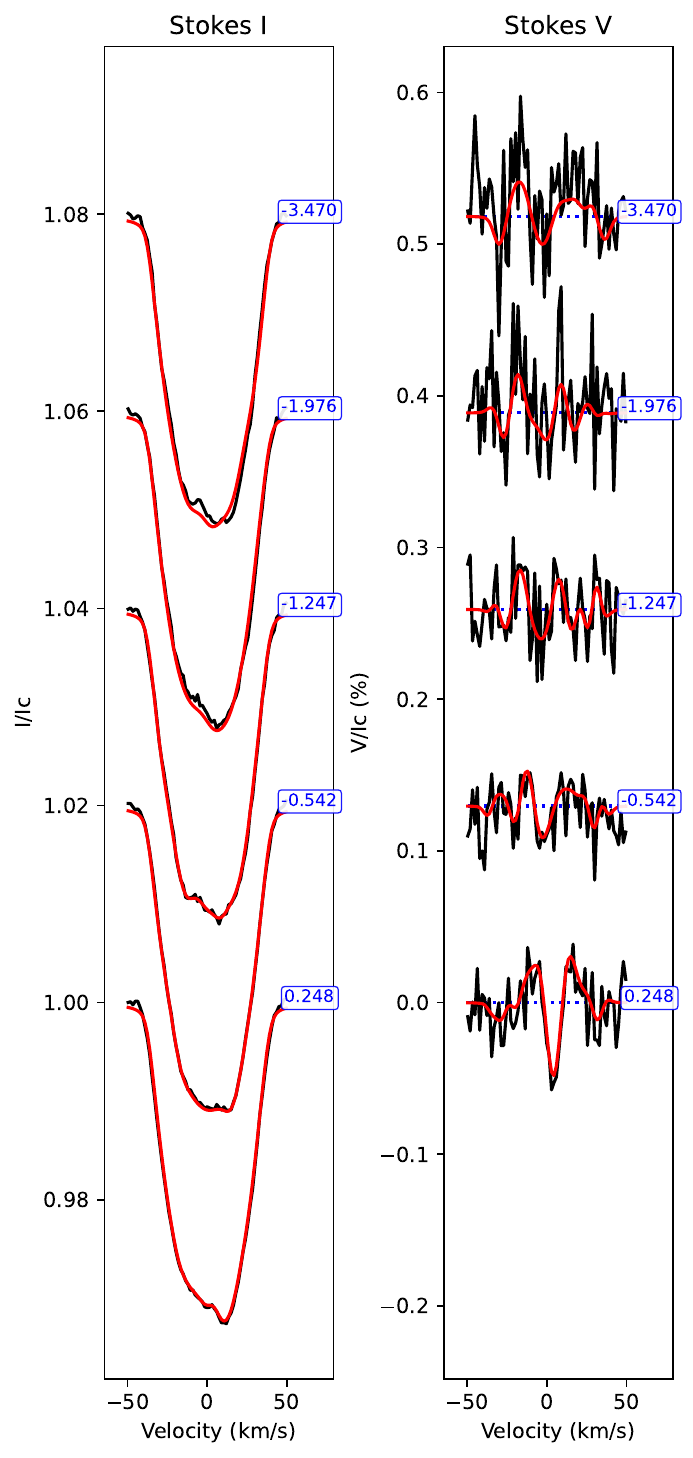} 
 \hspace{-3.6cm}
 \includegraphics[width=7.5cm]{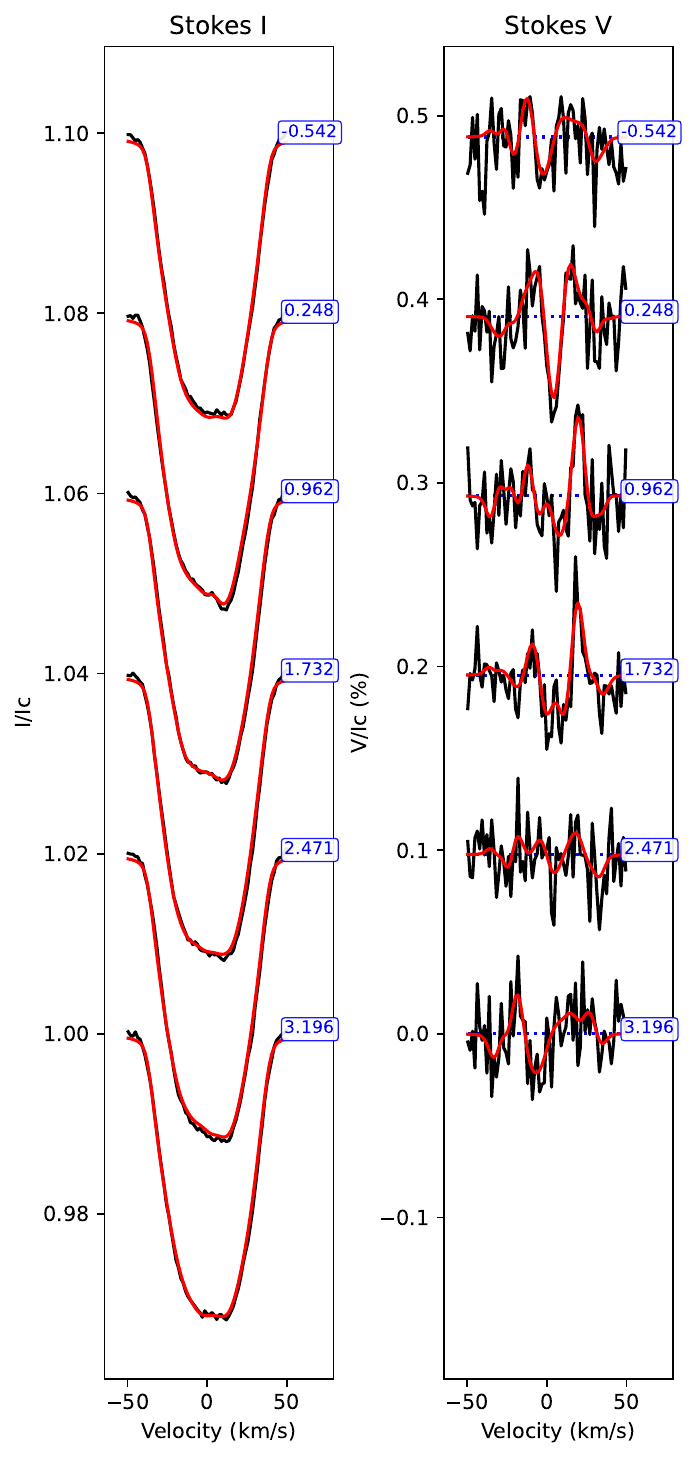}
 \hspace{-3.6cm}
 \includegraphics[width=7.5cm]{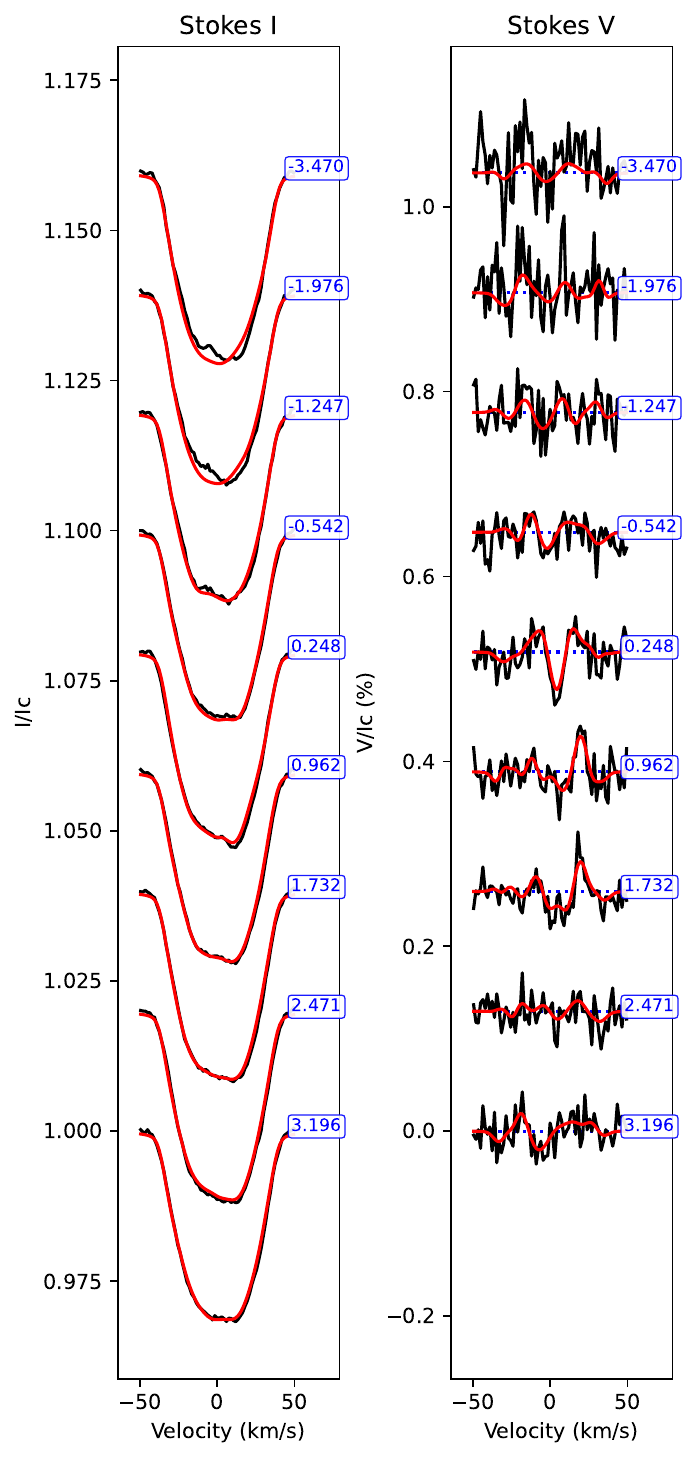}
\hspace{-4.3cm}
\vspace{0.5cm}
 \caption{Observations (black) and calculated (red) profiles for 2018 using the B24 rotation parameters. From left to right: Stokes I solutions using the first five, last six and all observations. Stokes V is only showed for the last case. The rotational phases are given for the stellar equator.}             
 \label{pr2018}
    \end{figure*}

The Stokes I and V spectra were modelled in the same way as described in \cite{Hackman2016}. This means that the intensity profile was calculated using an analytic approach, not by solving radiative transfer in stellar model atmospheres. This approximation is actually standard in ZDI analysis \citep[see, e.g.,][]{Petit2004b}. The continuum intensity $I_{\rm c}$ in Figs. \ref{pr2018} and \ref{profiles} and the brightness in the ZDI maps are for the wavelength region around the mean LSD wavelength, i.e. $\sim 5000$ \AA.

Since ZDI is a highly ill-posed inversion problem, regularisation has to be applied to the solution. For both the magnetic field and brightness, we applied the same regularisation as in the study by \cite{Rosen2016}. For the magnetic field this was based on penalising higher angular degrees and unnecessarily strong magnetic field by the function:

$$R_B = \sum_{\ell,m} \ell^2 (\alpha_{\ell,m}^2 + \beta_{\ell,m}^2 + \gamma_{\ell,m}^2),$$

\noindent where $\alpha_{\ell,m}$, $\beta_{\ell,m}$ and $\gamma_{\ell,m}$ are the spherical harmonic expansion coefficients, and $\ell$ and $m$ are the angular degree and azimuthal order.

For the brightness we used Tikhonov regularisation, with the addition of an extra penalty function $R_{\rm pen}$ to limit the solution to physically meaningful values. This procedure was also the same as explained in the paper by \cite{Rosen2016}. This additional function was

\begin{equation}
R_{\rm pen} = \left\{ \begin{array}{cl}
\sum_i C_1 (T_i - T_0)^2 & : \ T_i \leq T_0 \\
\sum_i C_2 (T_i - T_0)^2 & : \ T_i > T_0
\end{array} \right.
\label{pen}
\end{equation}

\noindent We used the values $C_1=1$, $C_2= 10$ and $T_0=1.1$.

The impact of all three regularisation/penalty functions was weighted by the coefficients $\Lambda_B$, $\Lambda_T$ and $\Lambda_{\rm pen}$. For the last one we used a constant value of $\Lambda_{\rm pen}=1 \cdot 10^{-6}$. This value was found by exploring parameters $\Lambda_{\rm pen} \in [10^{-4}, 10^{-8}]$ with logarithmic steps of $\Delta \lg \Lambda_{\rm pen}=-1$ and choosing the one that gave a good fit of the Stokes I observations with physically reasonable values in the brightness map, i.e., relative brightness values within the range $[0.2, 1.2]$.
The parameters $\Lambda_B$ and $\Lambda_T$ were adjusted during the inversion runs by studying the convergence of the deviation towards values corresponding to the $S/N$-ratios.

The stellar parameters of V889 Her are mostly well known through previous studies. The reported exceptionally high differential rotation had to be taken into account, since it may distort the surface pattern already during a few rotations.
In previous Doppler imaging and ZDI studies of magnetically active stars, a simple monotonic solar (or anti-solar) surface DR curve has been implemented. However, there is no solid basis for the assumption that this can generally be applied to other late-type stars. On the contrary, some studies show signs of deviation from this; see, e.g., Fig. 3 in  \cite{Kovari2015} and Fig. 8 in \cite{Kriskovics2023}. In addition, in numerical simulations, the resulting DR curve is generally more complex \citep[see, e.g.,][]{Viviani2018}.

To allow for a more complicated differential rotation, we
used the full solar differential rotation curve

\begin{equation}
\Omega(\varphi) = \Omega_{\mathrm{eq}}(1 + a \sin^2(\varphi) + b \sin^4(\varphi)),
\label{dr}
\end{equation}

\noindent where $\Omega(\varphi)$ and $\Omega_{\mathrm{eq}}$ are the angular velocity at latitude $\varphi$ and the equator, respectively. The coefficients $a$ and $b$ determine the amount and shape of the differential rotation curve.

Contrary to, e.g., the analysis by \cite{Brown2024}, we did not fit the differential rotation (DR) to the observations. Instead, we used three sets of rotation parameters published in recent studies of V889 Her. There were several reasons for this choice. Firstly, our rotation phase coverage, although enough for the ZDI analysis, was not deemed sufficient to constrain DR. 
Secondly, the approach to choose the rotation parameters that give the best fit to the data is not completely robust. It does not take into account that increasing the number of parameters will always give a better fit. This could be amended by using, e.g., the Bayesian Information Criterion (BIC). However, the use of regularisation will also influence the convergence of the ZDI solution, and this may cause a bias favoring certain combinations. This is even more complicated by possible artifacts, which usually appear as vertical or arch like structures in ZDI images \citep[see, e.g.,][]{Hackman2001}.

When searching for the combination of DR and the rotation period at the equator, $P_\mathrm{eq}$, there are two factors that will contribute to a better solution: The overall shape of the profile and the rotationally modulated signal. For the first factor, spots rotating with the stellar surface could contribute to a mean profile favoring a wrong DR, thus making the convergence of the inversion faster. As for the second factor, DR can only be detected if there are large spots at different latitudes. As this is usually not the case, there is often a degeneracy between $P_\mathrm{eq}$ and DR \citep[see, e.g., Fig. 2, ][]{Brown2024}.

Thus, we tested three sets of parameters used in previous studies: Rigid body rotation \citep[][W19]{Willamo2019}, solar-like monotonic DR \citep[][B24]{Brown2024} and non-monotonic DR
\citep[][T24]{Tuomi2024}. The set of parameters are listed in Table \ref{spar}.

Implementing DR means that we have to specify a "zero time" to which each image is "rewound", i.e., each latitude is shifted to compensate for the rotational shear. Thus, the rotational phases of the spectra reported in this study are for the equator. These were calculated using the ephemeris

$${\rm BJD}_{\phi =0} = {\rm BJD}_0 + P_{\rm eq} \cdot E,$$

\noindent where BJD$_0$ is given in Table \ref{obssum} and $P_{\rm eq}$ are listed in Table \ref{spar}.

In the Sun DR is taken into account using "synoptic" maps. For V889 Her, this is not an option because our observations cover several rotations. Thus, in Figs. \ref{ZDI2017}--\ref{ZDI2023} the rotational phases are only indicated for the images using solid body rotation. Furthermore, the rewinding of the images will be seen as characteristic patterns in the resulting ZDI maps.

\section{Results}

The ZDI maps are shown in Figs. \ref{ZDI2017}--\ref{ZDI2023}. All maps were calculated using a consistent procedure in testing different gradually decreasing regularisation parameters. In choosing the final values, the decrease in the deviation in relation to the mean $S/N$ for each set was a criterion. 
The combination $\Lambda_B = 3 \cdot 10^{-9}$ and $\Lambda_T = 1 \cdot 10^{-6}$ was found to be suitable for all data. This fact made comparison of the deviation of the solutions easier, as a lower regularisation parameter 
always leads to a smaller error between the data and model.

A good solution covering the full observation run was found in all cases, except the observations from 2018. Looking at the fits of the line profiles in Fig. \ref{pr2018}, it is clear that there is a discrepancy between the first and last nights. In order to rule out DR, we explored the rotation parameter space for a monotonic DR curve with $P_{\rm eq} \in [1.3,1.4]$ and $a \in [0.0, 0.2]$ (Eq. \ref{dr}). This did not improve the solution. Thus, we proceeded to step by step remove spectra from the beginning and the end of the observation run. Both visual inspection and deviation values favoured a division of this data set into two partly overlapping sets with the first five and last six spectra. These are labelled 2018F5 and 2018L6. The other sets are labelled according to the year of the observations. 

The mean deviation of the ZDI solutions compared with the observations is listed in Table \ref{zdires}. It is dominated by the deviation of the Stokes-I profile. The natural reason for this is that the unpolarised spectra are most sensitive to modelling errors. The Stokes-V profiles are differences of two oppositely polarised spectra, where modelling errors are mostly eliminated by subtraction.

\begin{table*}
  \caption{Characteristics of the ZDI solutions. The energy in the toroidal field is naturally $p_{\rm tor} = 100\% - p_{\rm pol}$ and non-axisymmetric field  $p_{\rm nax} = 100\% - p_{\rm axi}$.}
  \centering
\begin{tabular}{llllllll}
  \hline \hline

Set & Rot. par. & Dev.& $\langle|B| \rangle$& $p_{\rm pol}$ & $p_{\rm axi}$ & $p_{\ell =1}$ & Peak $\ell$\\
    &           & $[\%]$ & [G] & [\%]& $[\%]$ & $[\%]$ &  \\
\hline
2017 & W19   & 0.0498 & 66 & 68.3 & 24.5 & 19.6 & 1\\
       & B24   & 0.0467 & 67 & 72.0 & 29.0 & 13.3 & 2 \\
       & T24   & 0.0466 & 66 & 69.9 & 27.4  & 17.0 & 3\\
\hline
2018F5 & W19 & 0.0631 & 64 & 63.0 & 23.0 &  32.1 & 1\\
         & B24 & 0.0597 & 64 & 63.9 & 28.8 &  31.7 & 1 \\
         & T24 & 0.061  & 71 & 60.4 & 26.4 & 32.5 & 1\\
\hline
2018L6 & W19 & 0.0585 & 66 & 77.8 & 33.5 & 29.7 & 1 \\
         & B24 & 0.0583 & 74 & 65.8 & 30.2 & 25.5 & 1\\
         & T24 & 0.0607 & 64 & 74.2 & 24.8 & 27.8 & 1\\
\hline
2019 & W19   & 0.0748 & 97 & 53.8 & 42.2 & 15.1 & 3 \\
       & B24   & 0.0745 & 97 & 56.6 & 30.7 & 14.9 & 3 \\
       & T24   & 0.0791 & 99 & 55.7 & 29.0 & 18.6 & 3\\   
\hline
2021 & W19   & 0.0568 & 83 & 29.7 & 64.3 &  16.9 & 3 \\
       & B24   & 0.0575 & 87 & 39.2 & 54.6 & 20.3 & 3 \\
       & T24   & 0.0585 & 84 & 38.7 & 58.4 & 22.5 & 3 \\    
\hline
2023 & W19   & 0.0591 & 80 & 56.0 & 32.2 & 33.1 & 1 \\
       & B24   & 0.0622 & 80 & 62.2 & 22.0 & 27.0  & 1\\
       & T24   & 0.0619 & 79 & 56.8 & 25.1 &  34.4 & 1\\    
\hline
\end{tabular}
\label{zdires}
\end{table*}

All brightness maps are dominated by a high latitude large spot structure. The coolest part of this is slightly below the rotational pole. Some smaller spot structures are seen around the equator, but these may be artifacts. This possibility is supported by the tendency that they are strongest in images recovered with less phase coverage (the sets from 2017 and 2018). 

Some main characteristics of the ZDI magnetic field maps are listed in Table \ref{zdires}: the mean magnetic field strength ($\langle|B| \rangle$), the fractions of magnetic energy in the poloidal ($p_{\rm pol}$) and axisymmetric $p_{\rm axi}$ field components, the same for the angular degree $\ell=1$, and the $\ell$ with most magnetic energy. The axisymmetry was defined as the fraction of the magnetic field energy in azimuthal order $m=0$. Using the alternative definition of $m < \frac{\ell}{2}$ did not significantly change the result.

As can be seen in Table \ref{zdires}, the impact of different rotation parameters on the poloidality and axisymmetry of the magnetic field is small. This is also true for the magnetic energy distribution with $\ell$. For this reason, we only list the latter for inversions using the B24 parameters in Table \ref{ldist}.

\begin{table*}
  \caption{Magnetic field energy distribution $[\%]$ with $\ell$ for the ZDI maps using the rotation parameters B24 (solar type differential rotation).}
  \centering
\begin{tabular}{lrrrrrrrrrrr}
  \hline \hline
Set & $\ell=1$  & $\ell=2$ & $\ell=3$ & $\ell=4$ & $\ell=5$ & $\ell=6$ & $\ell=7$ & $\ell=8$ & $\ell=9$ & $\ell=10$ & $\ell > 10$ \\
\hline \hline
2017 & 17.0 &  15.8 &  18.0 & 13.7 &  8.8 &  7.4 &  3.8 &  2.9 &  4.9 &  1.4 &  6.3 \\
2018F5 &  32.5 &  10.0 &  14.7 & 8.0 &  7.0 &  5.6 &  7.5 &  4.2 &  2.4 &  1.8 &  6.5 \\
2018L6 &  27.8 &  11.7 &  4.5 & 10.2 &  12.9 &  9.2 &  8.2 & 2.7 &  3.4 &  1.5 &  7.8 \\
2019 &  18.6 &  14.7 &  19.5 & 15.6 &  10.4 &  6.3 &  4.9 & 4.2 &  2.2 &  1.0 &  2.6 \\
2021 &  22.5 &  7.4 &  23.2 & 18.9 &  10.5 &  6.9 &  3.8 & 1.7 &  1.0 &  1.2 &  2.6 \\
2023 &  34.4 &  14.0 &  6.2 & 10.0 &  7.4 &  10.2 &  8.0 & 2.8 &  2.4 &  1.1 &  3.5 \\

\hline
\end{tabular}
\label{ldist}
\end{table*}

\section{Discussion}

During the years 2017 -- 2023, we can see clear changes especially in the structure of the radial field. Our findings confirm the results obtained by \cite{Brown2024}. The polar radial field changes from a negative polarity (2017) to a mixed polarity (2018-9 and 2023), and even a possible reversal around 2021. However, a persistent dominant structure is the very strong azimuthal field at high latitude. 
The azimuthal component appears much more stable than the radial component, exhibiting no polarity reversals, only some moderate changes in the topology of the high-latitude magnetic ring. This behavior cannot be associated with a solar-like magnetic cycle. It is unlikely that the 
changes in the radial component throughout the years are associated with a long-term spot evolution.

Compared with the images presented by \cite{Brown2024}, we retrieve a weaker magnetic field. Since there is overlap in time between these two studies, this could be seen as a discrepancy. However, it is a well known fact that the strength of the retrieved magnetic field in ZDI studies is highly dependent on the S/N and spectral resolution, through their effect on the optimal regularisation parameter. Thus, the difference can be explained by, e.g., that the wavelength coverage of SOFIN is limited, leading to a smaller number of spectral absorption lines available for the LSD computation.

In general, we get the best solutions with the surface DR derived by \cite{Brown2024}. However, the differences in the goodness of the solutions are surprisingly small between the three parameter sets. In fact, all of them give the best solution for at least one set. This, again, confirms the findings by \cite{Brown2024}, where differences in DR were reported between the observation sets.

Using a significant surface DR will, of course, introduce some characteristic patterns in the brightness and magnetic field maps. As mentioned in Section \ref{ZDI}, this is because each image is "rewound" to a certain zero timing.

The mean magnetic field shows a growing trend from 2017--8 to 2019--23. This would at first glance be surprising, since the spot activity is in a declining phase (Fig. \ref{photom}).
However, the amount of magnetic field recovered with ZDI is highly dependent on the S/N ratio \citep{Hackman2024}. Furthermore, as shown by \cite{Kochukhov2020} the recovered surface magnetic field in ZDI maps is only a fraction of the total field. Also, the study of another young solar type star \object{LQ Hya} showed some lack of correlation between the strongest magnetic field and cool spots \citep{Kochukhov2023}.

The poloidal component of the large-scale magnetic field is mostly dominating, except for the the 2021 set. This set also shows the most axisymmetric field.

The star also shows signs of significant changes in the surface spot and magnetic field configuration within timescales of a week. These cannot be explained merely by DR. Instead, these changes will complicate the determination of the DR curve. This is not completely surprising, as similar findings have been reported for observations from 2011 \citep{Willamo2022,Brown2024} and 2018--9 \citep{Brown2024}.

\section{Conclusions}

The young solar analogue V889 Her shows signs of exceptionally high
differential rotation (DR), which may be variable and non-monotonic with respect to the stellar latitude.
In order to determine the shape of the DR curve, monitoring for longer time sequences than in this study would be needed. Ideally the spectropolarimetric observations should be accompanied by simultaneous space based high accuracy photometry.

The topology of the large-scale magnetic field is mostly rather constant, with strong poloidal and non-axisymmetric components. However, in 2021 the field is predominantly toroidal and axisymmetric. Regarding the field complexity, most magnetic energy is in low angular degrees, $\ell=1$ usually containing the peak. 

In June 2018 we see signs of quite abrupt changes in both the large scale surface magnetic field and spot distribution. These changes cannot be explained by mere DR. The time scale of the changes is around one week. Similar events have been reported earlier for this star.

The observations in this study were timed during an apparently monotonic decrease in the spot activity of V889 Her. Thus, the observed changes in the field topology cannot be associated with a possible stellar spot cycle.

\begin{acknowledgements}
Based on observations made with the Nordic Optical Telescope, owned in collaboration by the University of Turku and Aarhus University, and operated jointly by Aarhus University, the University of Turku and the University of Oslo, representing Denmark, Finland and Norway, the University of Iceland and Stockholm University at the Observatorio del Roque de los Muchachos, La Palma, Spain, of the Instituto de Astrofisica de Canarias. The NOT data were obtained under program IDs P51-502, P57-504, P63-502 and P67-504. This work was supported by the Research Council of Finland through funding of the consortium project PEAS (decision 376613).
O.K. acknowledges support by the Swedish Research Council (grant agreement no. 2023-03667) and the Swedish National Space Agency.
M.T. acknowledges the Jenny and Antti Wihuri foundation for financial support.
SW was supported by the Research Council of Norway
 through tthe Centres of Excellence scheme, 
project number 262622 (“Rosseland Centre for Solar Physics”).
\end{acknowledgements}

\bibliographystyle{aa}
\bibliography{hackman}

\begin{thebibliography}{26}
\expandafter\ifx\csname natexlab\endcsname\relax\def\natexlab#1{#1}\fi

\bibitem[{{Brown} {et~al.}(2024){Brown}, {Marsden}, {Jeffers}, {Heitzmann}, {Barnes}, \& {Folsom}}]{Brown2024}
{Brown}, E.~L., {Marsden}, S.~C., {Jeffers}, S.~V., {et~al.} 2024, \mnras, 528, 4092

\bibitem[{{Donati} {et~al.}(1997){Donati}, {Semel}, {Carter}, {Rees}, \& {Collier Cameron}}]{Donati1997}
{Donati}, J.~F., {Semel}, M., {Carter}, B.~D., {Rees}, D.~E., \& {Collier Cameron}, A. 1997, \mnras, 291, 658

\bibitem[{{Hackman} {et~al.}(2001){Hackman}, {Jetsu}, \& {Tuominen}}]{Hackman2001}
{Hackman}, T., {Jetsu}, L., \& {Tuominen}, I. 2001, \aap, 374, 171

\bibitem[{{Hackman} {et~al.}(2024){Hackman}, {Kochukhov}, {Viviani}, {Warnecke}, {Korpi-Lagg}, \& {Lehtinen}}]{Hackman2024}
{Hackman}, T., {Kochukhov}, O., {Viviani}, M., {et~al.} 2024, \aap, 682, A156

\bibitem[{{Hackman} {et~al.}(2016){Hackman}, {Lehtinen}, {Ros{\'e}n}, {Kochukhov}, \& {K{\"a}pyl{\"a}}}]{Hackman2016}
{Hackman}, T., {Lehtinen}, J., {Ros{\'e}n}, L., {Kochukhov}, O., \& {K{\"a}pyl{\"a}}, M.~J. 2016, \aap, 587, A28

\bibitem[{{Ilyin}(2012)}]{Ilyin2012}
{Ilyin}, I. 2012, Astronomische Nachrichten, 333, 213

\bibitem[{{Ilyin}(2000)}]{Ilyin2000}
{Ilyin}, I.~V. 2000, PhD thesis, University of Oulu, Division of Astronomy

\bibitem[{{Kazarovets} \& {Samus}(1997)}]{Kazarovets1997}
{Kazarovets}, E.~V. \& {Samus}, N.~N. 1997, Information Bulletin on Variable Stars, 4471, 1

\bibitem[{{K{\H{o}}v{\'a}ri} {et~al.}(2015){K{\H{o}}v{\'a}ri}, {Kriskovics}, {K{\"u}nstler}, {Carroll}, {Strassmeier}, {Vida}, {Ol{\'a}h}, {Bartus}, \& {Weber}}]{Kovari2015}
{K{\H{o}}v{\'a}ri}, Z., {Kriskovics}, L., {K{\"u}nstler}, A., {et~al.} 2015, \aap, 573, A98

\bibitem[{{Kochukhov}(2026)}]{Kochukhov2026}
{Kochukhov}, O. 2026, \aap, 707, A16

\bibitem[{{Kochukhov} {et~al.}(2023){Kochukhov}, {Hackman}, \& {Lehtinen}}]{Kochukhov2023}
{Kochukhov}, O., {Hackman}, T., \& {Lehtinen}, J.~J. 2023, \aap, 680, L17

\bibitem[{{Kochukhov} {et~al.}(2020){Kochukhov}, {Hackman}, {Lehtinen}, \& {Wehrhahn}}]{Kochukhov2020}
{Kochukhov}, O., {Hackman}, T., {Lehtinen}, J.~J., \& {Wehrhahn}, A. 2020, \aap, 635, A142

\bibitem[{{Kochukhov} {et~al.}(2014){Kochukhov}, {L{\"u}ftinger}, {Neiner}, {Alecian}, \& {MiMeS Collaboration}}]{Kochukhov2014}
{Kochukhov}, O., {L{\"u}ftinger}, T., {Neiner}, C., {Alecian}, E., \& {MiMeS Collaboration}. 2014, \aap, 565, A83

\bibitem[{{Kochukhov} {et~al.}(2010){Kochukhov}, {Makaganiuk}, \& {Piskunov}}]{Kochukhov2010}
{Kochukhov}, O., {Makaganiuk}, V., \& {Piskunov}, N. 2010, \aap, 524, A5

\bibitem[{{Kriskovics} {et~al.}(2023){Kriskovics}, {K{\H{o}}v{\'a}ri}, {Seli}, {Ol{\'a}h}, {Vida}, {Henry}, {Granzer}, \& {G{\"o}rgei}}]{Kriskovics2023}
{Kriskovics}, L., {K{\H{o}}v{\'a}ri}, Z., {Seli}, B., {et~al.} 2023, \aap, 674, A143

\bibitem[{{Lehtinen} {et~al.}(2011){Lehtinen}, {Jetsu}, {Hackman}, {Kajatkari}, \& {Henry}}]{Lehtinen2011}
{Lehtinen}, J., {Jetsu}, L., {Hackman}, T., {Kajatkari}, P., \& {Henry}, G.~W. 2011, \aap, 527, A136

\bibitem[{{Petit} {et~al.}(2004){Petit}, {Donati}, {Wade}, {Landstreet}, {Bagnulo}, {L{\"u}ftinger}, {Sigut}, {Shorlin}, {Strasser}, {Auri{\`e}re}, \& {Oliveira}}]{Petit2004b}
{Petit}, P., {Donati}, J.-F., {Wade}, G.~A., {et~al.} 2004, \mnras, 348, 1175

\bibitem[{{Piskunov} {et~al.}(1995){Piskunov}, {Kupka}, {Ryabchikova}, {Weiss}, \& {Jeffery}}]{VALD1}
{Piskunov}, N.~E., {Kupka}, F., {Ryabchikova}, T.~A., {Weiss}, W.~W., \& {Jeffery}, C.~S. 1995, \aaps, 112, 525

\bibitem[{{Ros\'en} {et~al.}(2016){Ros\'en}, {Kochukhov}, {Hackman}, \& {Lehtinen}}]{Rosen2016}
{Ros\'en}, L., {Kochukhov}, O., {Hackman}, T., \& {Lehtinen}, J. 2016, \aap, 593, A35

\bibitem[{{Ryabchikova} {et~al.}(2015){Ryabchikova}, {Piskunov}, {Kurucz}, {Stempels}, {Heiter}, {Pakhomov}, \& {Barklem}}]{VALD2}
{Ryabchikova}, T., {Piskunov}, N., {Kurucz}, R.~L., {et~al.} 2015, \physscr, 90, 054005

\bibitem[{{Strassmeier} {et~al.}(2003){Strassmeier}, {Pichler}, {Weber}, \& {Granzer}}]{Strassmeier2003}
{Strassmeier}, K.~G., {Pichler}, T., {Weber}, M., \& {Granzer}, T. 2003, \aap, 411, 595

\bibitem[{{Tuomi} {et~al.}(2024){Tuomi}, {Lehtinen}, {Henry}, \& {Hackman}}]{Tuomi2024}
{Tuomi}, M., {Lehtinen}, J.~J., {Henry}, G.~W., \& {Hackman}, T. 2024, \aap, 689, A262

\bibitem[{{Viviani} {et~al.}(2018){Viviani}, {Warnecke}, {K{\"a}pyl{\"a}}, {K{\"a}pyl{\"a}}, {Olspert}, {Cole-Kodikara}, {Lehtinen}, \& {Brandenburg}}]{Viviani2018}
{Viviani}, M., {Warnecke}, J., {K{\"a}pyl{\"a}}, M.~J., {et~al.} 2018, \aap, 616, A160

\bibitem[{{Willamo} {et~al.}(2019){Willamo}, {Hackman}, {Lehtinen}, {K{\"a}pyl{\"a}}, {Ilyin}, {Henry}, {Jetsu}, {Kochukhov}, \& {Piskunov}}]{Willamo2019}
{Willamo}, T., {Hackman}, T., {Lehtinen}, J.~J., {et~al.} 2019, \aap, 622, A170

\bibitem[{{Willamo} {et~al.}(2022{\natexlab{a}}){Willamo}, {Hackman}, {Lehtinen}, {Korpi-Lagg}, \& {Kochukhov}}]{Willamo2022}
{Willamo}, T., {Hackman}, T., {Lehtinen}, J.~J., {Korpi-Lagg}, M., \& {Kochukhov}, O. 2022{\natexlab{a}}, The Open Journal of Astrophysics, 5, 10

\bibitem[{{Willamo} {et~al.}(2022{\natexlab{b}}){Willamo}, {Lehtinen}, {Hackman}, {K{\"a}pyl{\"a}}, {Kochukhov}, {Jeffers}, {Korhonen}, \& {Marsden}}]{Willamo2022a}
{Willamo}, T., {Lehtinen}, J.~J., {Hackman}, T., {et~al.} 2022{\natexlab{b}}, \aap, 659, A71

\end{thebibliography}

%

\begin{appendix}

\onecolumn

\section{Observations}

In Table \ref{obs} the dates, BJD:s, exposure times including all subexposures and S/N measured at $\lambda = 6302.0$ \AA~ is given.

\begin{table}[H]
  \caption{The SOFIN observations of V889 Her 2017-23.}
  \centering
\begin{tabular}{ccccc}
  \hline \hline
Observing run & Date [DD/MM/YYYY] & BJD - 2400000& Exposure [hh:mm] & S/N \\
\hline
6/2017 & 04/06/2017 & 57908.615 & 00:40 & 259 \\
      & 09/06/2017 & 57913.558 &  00:40 & 299 \\
      & 10/06/2017 & 57914.561 & 00:40 & 281 \\
      & 11/06/2017 & 57915.562 & 00:40 & 255 \\
      & 13/06/2017 & 57917.587 &  00:52 & 314 \\
      & 14/06/2017 & 57918.522 & 00:40 & 291 \\
\hline
6/2018 & 15/06/2018 & 58284.623 & 01:00 & 258 \\
      & 17/06/2018 & 58286.607 &  00:52 & 240 \\
      & 18/06/2018 & 58287.575 &  00:52 & 274 \\
      & 18/06/2018 & 58288.514 &  01:31 & 334 \\
      & 20/06/2018 & 58289.563 &  01:31 & 328 \\
      & 20/06/2018 & 58290.509 &  01:18 & 331 \\
      & 22/06/2018 & 58291.531 & 00:52 & 327 \\
      & 22/06/2018 & 58292.513 & 00:52 & 328 \\
      & 23/06/2018 & 58293.475 & 00:52 & 324 \\
\hline
6/2019 & 20/06/2019 & 58655.470 & 00:52 & 324 \\
      & 21/06/2019 & 58655.706 & 00:40 & 293 \\
      & 22/06/2019 & 58656.545 & 00:52 & 329 \\
      & 23/06/2019 & 58657.531 & 00:52 & 340 \\
      & 24/06/2019 & 58658.534 & 00:52 & 316 \\
      & 25/06/2019 & 58659.609 & 00:52 & 339 \\
      & 25/06/2019 & 58660.467 & 00:52 & 343 \\
      & 26/06/2019 & 58660.644 & 00:52 & 336 \\
      & 26/06/2019 & 58661.468 & 00:52 & 338 \\
      & 27/06/2019 & 58662.505 & 00:52 & 346 \\
      & 28/06/2019 & 58662.705 & 00:40 & 297 \\
      & 29/06/2019 & 58664.463 & 00:52 & 356 \\
\hline
7/2021 & 17/07/2021 & 59412.541 & 00:52 & 294 \\
      & 17/07/2021 & 59413.501 & 01:18 & 285 \\
      & 19/07/2021 & 59414.548 & 01:31 & 337 \\
      & 19/07/2021 & 59415.504 & 00:52 & 335 \\
      & 20/07/2021 & 59415.636 & 00:52 & 316 \\
      & 20/07/2021 & 59416.456 & 00:52 & 354 \\
      & 21/07/2021 & 59417.424 & 00:52 & 340 \\
      & 22/07/2021 & 59418.421 & 00:52 & 281 \\
      & 23/07/2021 & 59419.424 & 01:57 & 287 \\
\hline
7/2023 & 27/06/2023 & 60123.468 & 00:52 & 199 \\
      & 28/06/2023 & 60124.468 & 00:52 & 360 \\
      & 29/06/2023 & 60125.464 & 00:52 & 318 \\
      & 30/06/2023 & 60126.464 & 00:52 & 333 \\
      & 01/07/2023 & 60126.608 & 00:52 & 295 \\
      & 01/07/2023 & 60127.461 & 00:52 & 287 \\
      & 02/07/2023 & 60128.467 & 00:52 & 326 \\
      & 04/07/2023 & 60129.547 & 00:52 & 311 \\
      & 05/07/2023 & 60130.564 & 00:52 & 361 \\
      & 06/07/2023 & 60131.620 & 00:52 & 336 \\
      & 07/07/2023 & 60132.625 & 00:52 & 341 \\
\hline
\hline
\end{tabular}
\label{obs}
\end{table}

\newpage

\section{Observed and calculated spectra for the stellar rotation parameters W19}
\label{appB}

\begin{figure*}[h]
 \includegraphics[width=6cm, height=9.5cm]{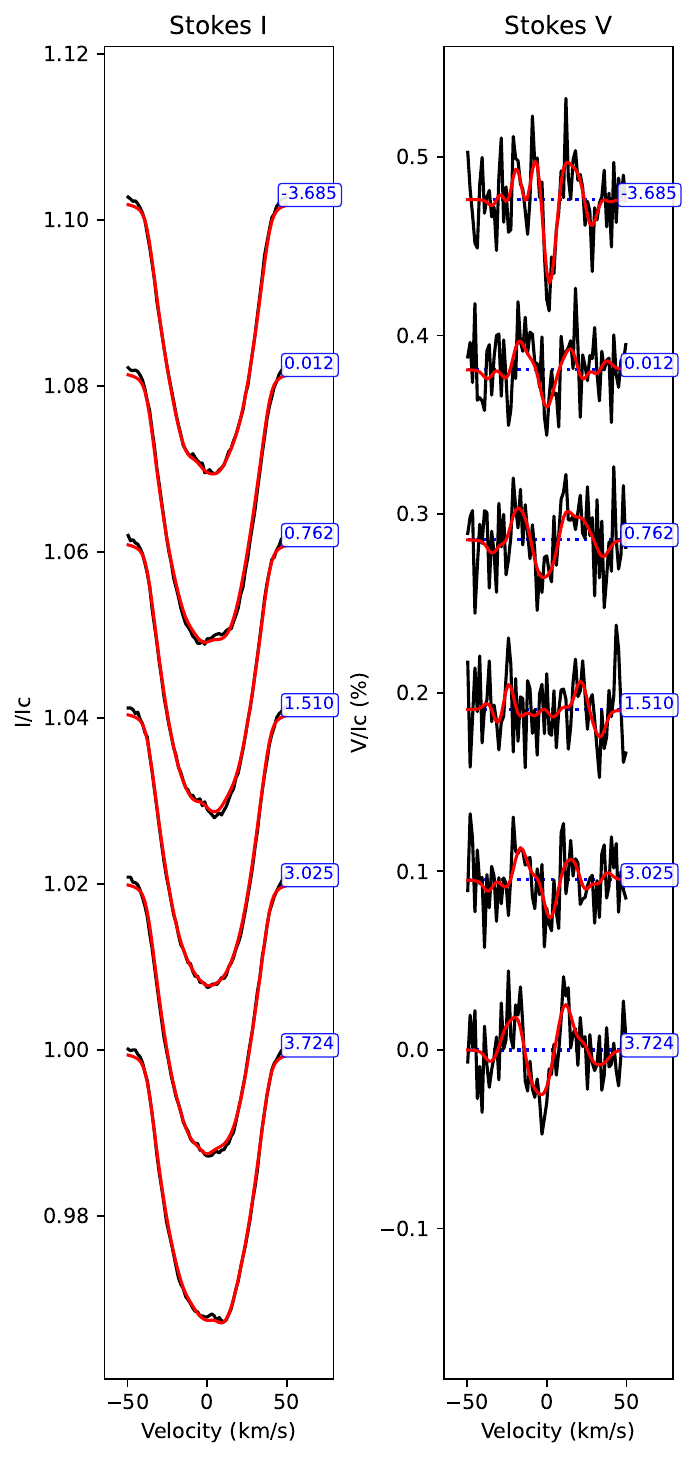} 
 \includegraphics[width=6cm, height=9.5cm]{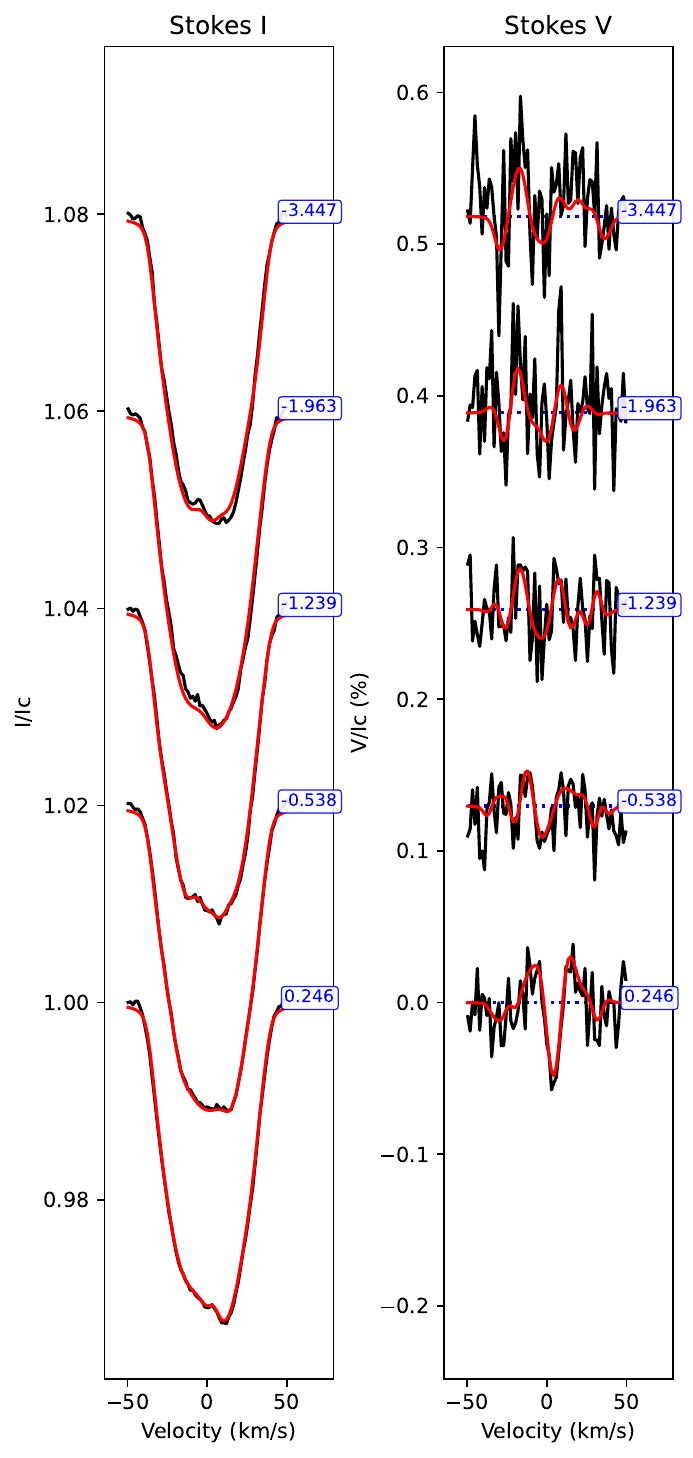}
 \includegraphics[width=6cm, height=9.5cm]{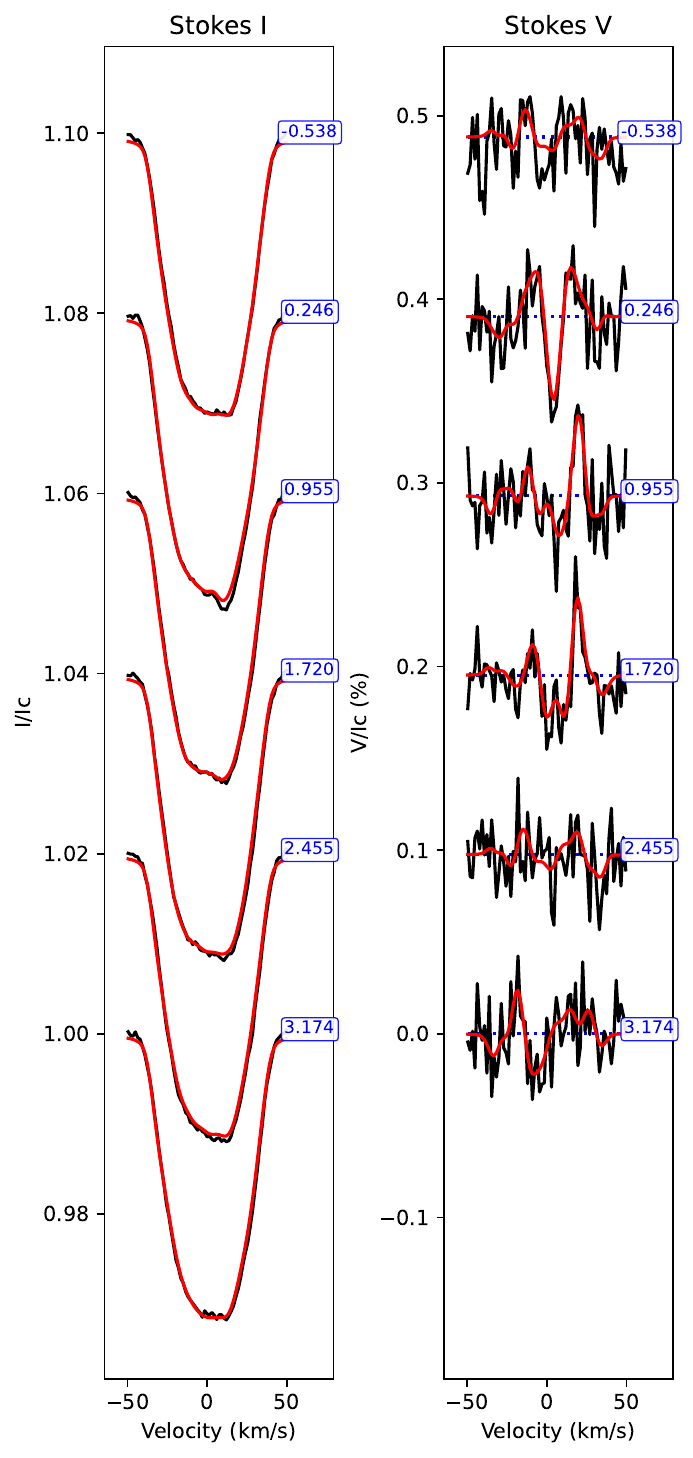}

 \includegraphics[width=6cm, height=9.5cm]{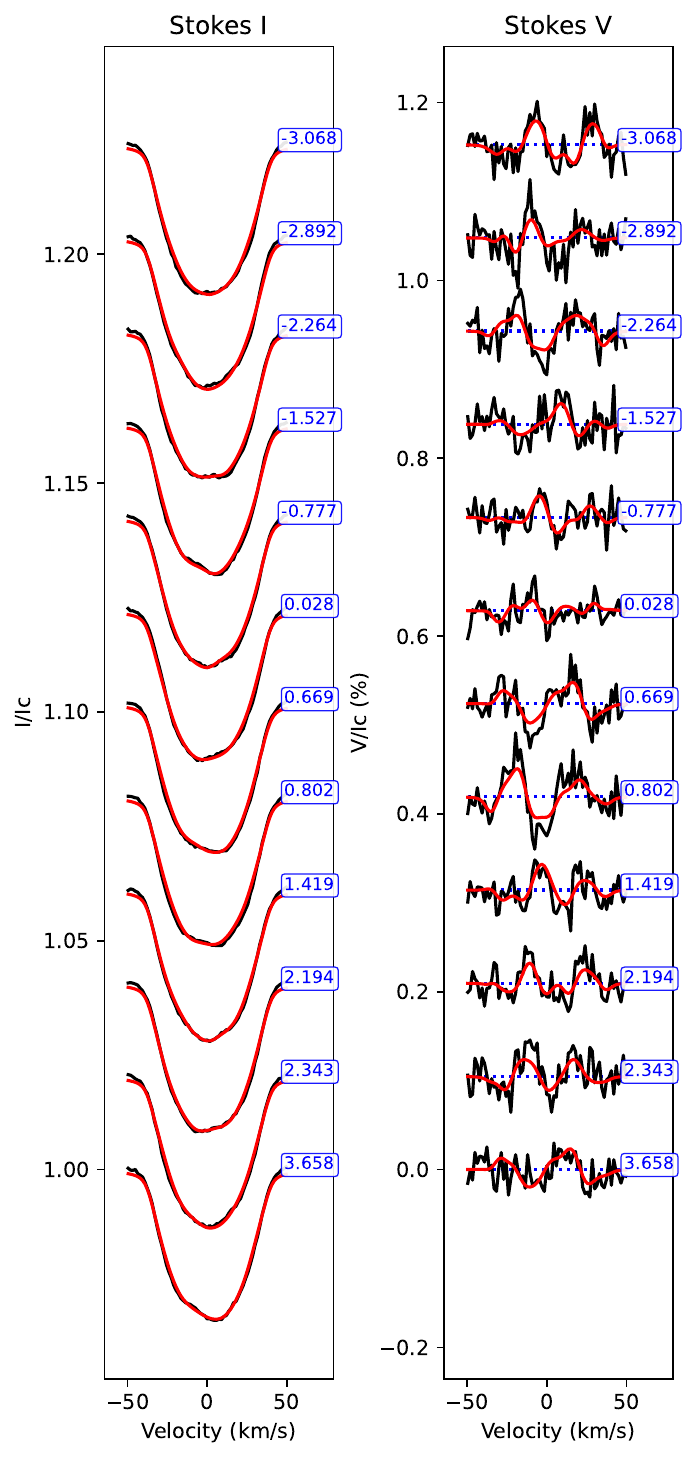}
 \includegraphics[width=6cm, height=9.5cm]{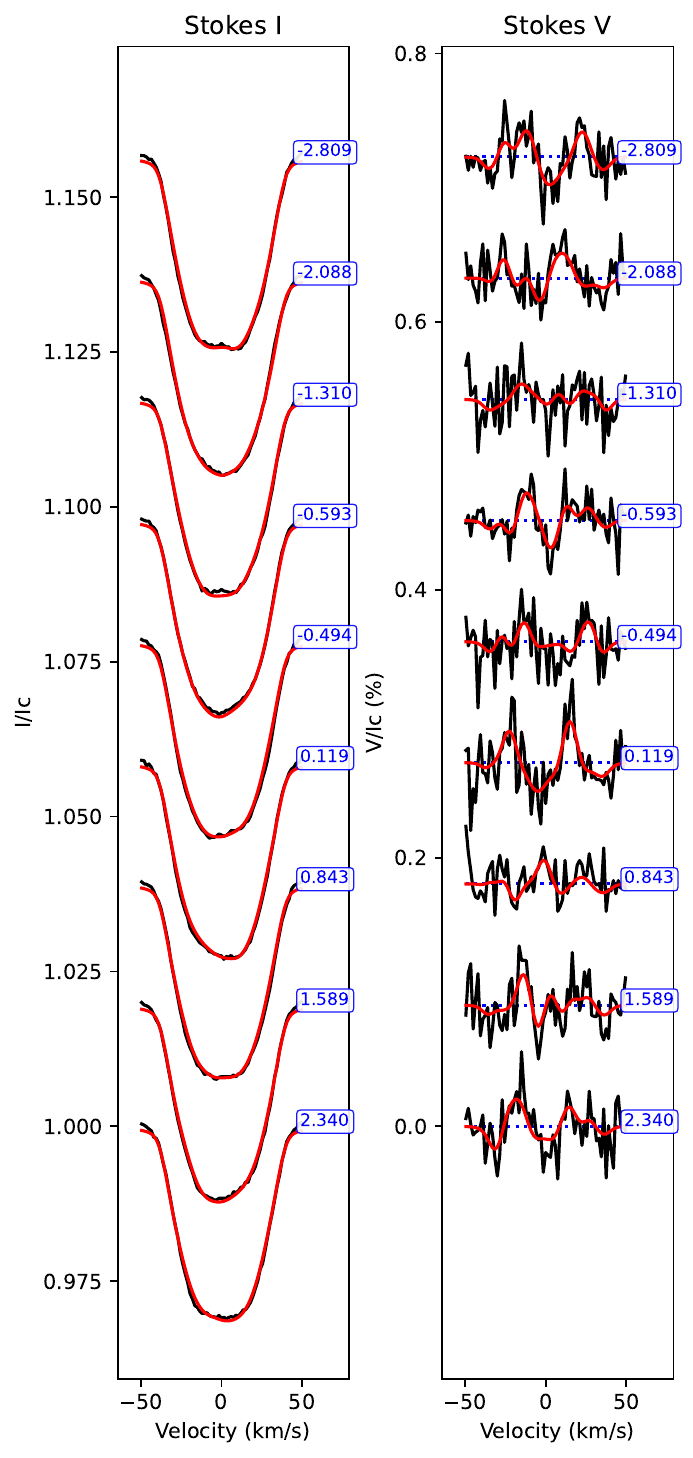}
 \includegraphics[width=6cm, height=9.5cm]{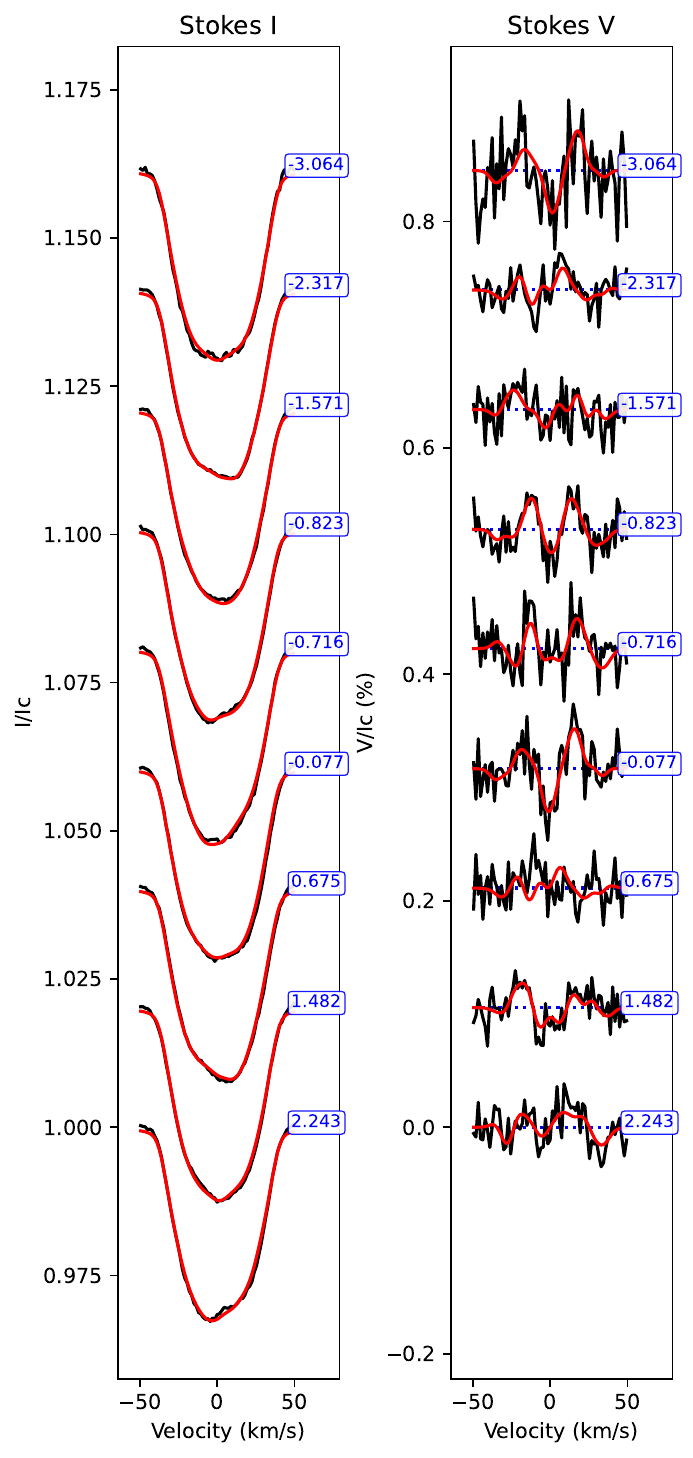}

\vspace{0.5cm}
 \caption{Observations (black) and calculated (red) profiles for the W19 rotation parameters. The sets are from left to right, top to bottom: 2017, 2018 first five, 2018 last six, 2019, 2021 and 2023.}
             
 \label{profiles}
    \end{figure*}

\end{appendix}

\end{document}